\documentclass[aps,prl,superscriptaddress,twocolumn,showpacs]{revtex4-1}
\usepackage{amsmath}
\usepackage{amssymb}
\usepackage{amsfonts}
\usepackage[normalem]{ulem}
\usepackage{subfigure}
\usepackage{graphicx}
\usepackage{dcolumn}
\usepackage{bm}
\usepackage{hyperref}
\usepackage{color}
\usepackage[english]{babel}

\begin{document}
	\title{ Electronic Band Structure and Superconducting Properties of SnAs }

	\author{P. I.~Bezotosnyi }
		\email{bezpi@sci.lebedev.ru}
		\affiliation{P. N. Lebedev Physical Institute, Russian Academy of Sciences, Moscow, 119991, Russia}
	\author{K.~A.~Dmitrieva}%
		\affiliation{P. N. Lebedev Physical Institute, Russian Academy of Sciences, Moscow, 119991, Russia}
	\author{A.~V.~Sadakov}
		\affiliation{P. N. Lebedev Physical Institute, Russian Academy of Sciences, Moscow, 119991, Russia}
	\author{K.~S.~Pervakov}
		\affiliation{P. N. Lebedev Physical Institute, Russian Academy of Sciences, Moscow, 119991, Russia}
	\author{A.~V.~Muratov}
		\affiliation{P. N. Lebedev Physical Institute, Russian Academy of Sciences, Moscow, 119991, Russia}
	\author{A.~S.~Usoltsev}
		\affiliation{P. N. Lebedev Physical Institute, Russian Academy of Sciences, Moscow, 119991, Russia}
		\affiliation{National Research University Higher School of Economics,  Moscow 101000, Russia}
	\author{A.~Yu.~Tsvetkov}
		\affiliation{P. N. Lebedev Physical Institute, Russian Academy of Sciences, Moscow, 119991, Russia}
	\author{S.~Yu.~Gavrilkin}
		\affiliation{P. N. Lebedev Physical Institute, Russian Academy of Sciences, Moscow, 119991, Russia}
	\author{N.~S.~Pavlov}%
		\affiliation{Institute for Electrophysics, Russian Academy of Sciences, Ural Branch, Ekaterinburg, 620016, Russia}
	\author{A.~A.~Slobodchikov}%
		\affiliation{Institute for Electrophysics, Russian Academy of Sciences, Ural Branch, Ekaterinburg, 620016, Russia}
		\affiliation{Kirensky Institute of Physics, Federal Research Center KSC SB RAS, Krasnoyarsk, 660036,  Russia}
	\author{O.~Yu.~Vilkov}
		\affiliation{Saint Petersburg State University, Saint Petersburg, 198504, Russia}
	\author{A.~G.~Rybkin}
		\affiliation{Saint Petersburg State University, Saint Petersburg, 198504, Russia}
	\author{I.~A. Nekrasov}%
		\affiliation{Institute for Electrophysics, Russian Academy of Sciences, Ural Branch, Ekaterinburg, 620016, Russia}
	\author{V.~M.~Pudalov}
		\affiliation{P. N. Lebedev Physical Institute, Russian Academy of Sciences, Moscow, 119991, Russia}
		\affiliation{National Research University Higher School of Economics,  Moscow 101000, Russia}


	\begin{abstract}

		We report comprehensive study of physical properties of the binary superconductor compound SnAs.
		The electronic band structure of SnAs was investigated using both angle-resolved photoemission spectroscopy (ARPES) in a wide binding energy range and density functional theory (DFT) within generalized gradient approximation (GGA). 
		The DFT/GGA calculations were done including spin-orbit coupling for both bulk and 
		(111) slab crystal structures. 
		Comparison of the DFT/GGA band dispersions with ARPES data shows that (111) slab much better describes ARPES data than just bulk bands. 
		Superconducting properties of SnAs were studied experimentally by specific heat, magnetic susceptibility,  magnetotransport measurements and Andreev reflection spectroscopy. 
		Temperature dependences of the superconducting gap  and of the specific heat were found to be well consistent with those expected for the single band BCS superconductors with an isotropic s-wave order parameter.  
		Despite spin-orbit coupling is present in SnAs, our data shows no signatures of a potential unconventional superconductivity, and the characteristic BCS ratio $2\Delta/T_c = 3.48 - 3.73$ is very close to the BCS value in the weak coupling limit. 
		
	\end{abstract}

	\maketitle


	\section{\label{sec:level1}Introduction}
	
		Binary compounds  of the SnX-family (X=Te, Se, As, S, Sb, P) attract much attention due to their unique properties \cite{Zhao2014, Hsieh2012, Xia2016, Sun2013, Kamitani2017}. 
		The most extensively studied SnSe \cite{Wang2015}  and SnTe \cite{Tanaka2012,  Tanaka2013} have  NaCl-structure. 
		These compounds were theoretically and experimentally determined as topological crystalline insulators. 
		In this class of materials topological properties are protected by crystal symmetries, that differs from conventional topological  insulators, where the time-reversal symmetry is  the determining factor   \cite{Fu2011}. 
		Moreover,  In-doped SnTe was found to manifest  signatures of Andreev bound states \cite{Sasaki2012}, which are characteristic of unconventional superconductivity. 
		Given this fact and the topological nature of the material,  Sn$_{1-x}$In$_{x}$Te is considered as a strong candidate for a topological superconductor \cite{Sato2017, Sato2013}.
		
		Tin arsenide is isostructural to SnTe and SnSe. 
		Without doping, this compound demonstrates superconducting properties, which were first reported in 1964 by Geller and Hull \cite{Geller1964}. 
		For a long time there was an uncertainty about valence state of Sn in this compound and about its possible influence on superconductivity. 
		Later,  Wang et. al \cite{Wang2014} experimentally demonstrated that Sn has a single valence state Sn$^{+3}$. 
		Also, from rough estimates based on experimental data it was  suggested, that SnAs is   a type-I  superconductor  with weak-coupling. 
		Temperature dependences of the critical magnetic field $H_{\rm c}$ and of the specific heat in zero   field are well described by the BCS model. 
		Nevertheless,  up to the best of our knowledge no information  is available on the specific heat behavior  in non-zero magnetic fields and  on the temperature dependence of the superconducting gap. 
		Furthermore, from earlier theoretical investigation  of electronic band structure, electron-phonon interaction, and  superconductivity performed in \cite{Tutuncu2015}, the calculated electron-phonon coupling parameter seemed to agree  with the one estimated  from specific heat data \cite{Wang2014}, however there is a  discrepancy between the theoretical and experimental values of T$_c$.
		
		Another interesting result   was found in investigations under high-pressure. 
		In Ref.~\cite{SREENIVASAREDDY2017}  SnAs was shown to exhibit a structural phase transition from NaCl- to CsCl-structure at around 37\,GPa, which confirms previously obtained experimental results \cite{Demishev1989}.   
		Moreover, there is  dramatic increase of $T_{\rm c}$ up to 12.2\,K at a critical point of the structural transition.
		
		On the theoretical side, in Ref.~\cite{HASE2016}  electronic band structure of SnAs   was calculated within the tight-binding approximation. 
		Despite the experimental data of Ref.~\cite{Wang2014}  demonstrated that Sn has a
		single valence state Sn$^{+3}$, this compound is considered \cite{HASE2016} as a potential candidate  for the valence-skip material. 
		The valence skipping  may lead to a negative effective Coulomb interaction between electrons. 
		The calculation results  reveal that the Sn$^x$ state in SnAs  is likely intermediate between valence skipper and fixed valence compound;  as a result, moderate charge fluctuations (and electron-phonon interaction) can be responsible for superconductivity. 
		
		The band dispersion of SnAs resembles that for SnTe, the known crystalline topological insulator. 
		Moreover, due to the similarity of SnAs to SnTe, it was suggested, that the mixed compound Sn(As,Te) may manifest topological superconductivity, if it is superconducting. 
		Preliminary experimental studies of  electronic band structure of SnAs by  ARPES  technique \cite{Bezotosnyi2017} have shown the presence of features that can be interpreted as a manifestation of the spin-orbit coupling (SOC): energy band splitting in the vicinity of the $\Gamma$ point, which  might be  important in context of possible topological nature of the mixed compound Sn(As, Te).  
		The ARPES data \cite{Bezotosnyi2017} was obtained for the (111) sample surface, whereas band structure calculations are available for bulk dispersion solely. 
		This impedes comparison of the data with theory; obviously, one needs band structure calculated within the corresponding cross-section of the Brillouin zone.
		
		Despite the existence of previous investigations, there is still lack of knowledge of the origin of  superconductivity in SnAs, and the question about the impact of SOC is still opened. 
		In the  current paper we  report detailed experimental study of the normal and superconducting properties of  SnAs. 
		For the normal state we studied by ARPES technique electronic band structure in a wide binding energy interval. 
		The measured superconducting properties include specific heat and magnetic susceptibility  in various magnetic felds, and  the superconducting gap by Andreev reflection spectroscopy. 
		Also, we report refined calculations of the SnAs band structure within DFT/GGA for bulk  and (111) slab crystal structures, with and without SOC, and compare them with the ARPES spectra.

	\section{\label{sec:Exp}SAMPLES and EXPERIMENTAL DETAILS}

		\begin{figure}[h]
			\centering
			\includegraphics[width=8cm]{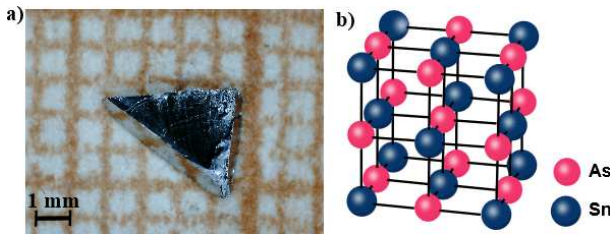} 	
			\caption{a)  Image of the  SnAs crystal.  b) Crystal structure of SnAs.}
			\label{fig:Sample}
		\end{figure}
		
		The SnAs samples were synthesized from pure elements: Sn and As with purity of 99.99\% and 99.9999\%, respectively. 
		The sealed quartz ampoule with elements taken in a stoichiometric ratio of 1:1 was slowly heated to $600^{\circ}$C, held for 48 hours to form the phase, then heated to $800^{\circ}$C and held for 24 hours to homogenize the melt. 
		The crystals were then grown  by a modified Bridgman method at a melt cooling rate of $1^{\circ}$C/hour down to 550$^{\circ}$C. 
		The grown crystals were  further annealed at a temperature of $550^{\circ}$C for 24 hours to remove growth defects.
		
		\begin{figure}[h]
			\centering
			\includegraphics[width=8cm]{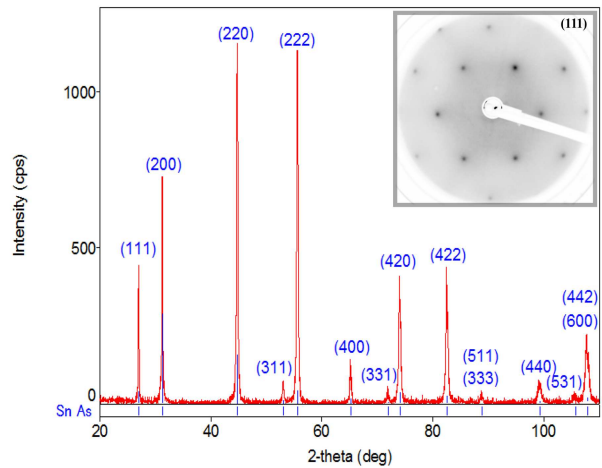} 	
			\caption{XRD spectrum of the SnAs (sample No.\,2). The abscissa represents the angle 2$\Theta$ in degrees, the ordinate is the reflection intensity. 	{\itshape{Inset}}: LEED pattern for SnAs crystal with electron beam energy $E=120$\,eV.}
			\label{fig:Structure}
		\end{figure}
	
		We studied two crystals (No.\,1 and No.\,2) obtained in two different growths. 
		Being nominally identical, the samples however had slightly different $T_c$ values. 
		For ARPES measurements both SnAs samples were used, for resistivity and Andreev spectroscopy we used sample No. 1, for specific heat and magnetic measurements -- sample No. 2.
		
		The obtained  SnAs crystals were cleaved into the small  pieces  of about 3\,mm size, which had a pyramidal shape with triangular base  (Fig.~\ref{fig:Sample}(a)). 
		Structural investigations of both SnAs crystals were done with Rigaku MiniFlex 600 X-ray powder diffractometer. 
		The measured diffraction pattern of SnAs is shown in Fig.~\ref{fig:Structure}. 
		It contains only reflexes related to the SnAs phase, and no reflections intrinsic to other phases. 
		From XRD analysis  we confirmed the NaCl-type structure (Fig. \ref{fig:Sample}(b)) and found  the lattice parameter $a = 5.723$\,\AA\,to be in agreement with earlier reported data \cite{Wang2014}. 
		The quality of the sample surface was investigated {\it in situ} by LEED (low--energy electron diffraction). 
		A regular hexagon with zero reflection at the center in the LEED pattern is shown  in Fig.~\ref{fig:Structure} (inset). 
		Since SnAs has the NaCl structure, the obtained pattern of reflections corresponds to the (111) plane. 
		Clear reflection pattern  also proves the single phase composition at the investigated surface. 
		
		Elemental composition of the sample was measured by EDS (Energy--Dispersive X-ray Spectroscopy) technique using SEM; the spectrum contains only lines related to the Sn- and As- atoms. 
		Quantitative analysis shows Sn and As content of 48.9\% and 51.1\%, respectively, that corresponds to the ratio of 0.978 : 1.022. 
		
		ARPES  measurements were performed using a semi-spherical VG Scienta R4000 energy analyzer with energy resolution of 17\,meV and angular resolution of $0.5^{\circ}$. 
		The ultraviolet helium lamp VUV 5k was used as a light source with photon energy $21.2$\,eV. The measurements were carried out at room temperature. 
		Pressure in the chamber didn't exceed $1-2 \times 10^{-10}$\,mbar.	
		The  samples  were cleaved along the (111) plane (see the LEED pattern on Fig.~\ref{fig:Structure}, inset, as described above). 
		Before measuring photoelectron spectra, the surface of the samples was cleaned by  Ar-ion beam and  subsequent heating at temperature of $150^{\circ}$C under ultrahigh  vacuum.
			
		Resistivity, magnetic and thermal properties of SnAs sample were measured using PPMS-9 (Quantum Design). 	
		Specific heat  measurements of SnAs crystals were carried out by thermal relaxation technique  with $^3$He calorimeter,  in the range of temperatures 0.4 -- 6\,K and magnetic fields 0 -- 500\,Oe. 
		Low-frequency magnetic susceptibility measurements of SnAs crystals were taken   in the range of temperatures 2 -- 6\,K  and  magnetic fields 0 -- 200\,Oe. 
		It is worth of noting, that magnetic susceptibility  at ``H=0" is actually measured at a finite low value (2--5\,Oe) of the AC-magnetic field while the external DC-magnetic field is set to zero. 
		Magnetization loop was measured using Vibrating Sample Magnetometer (VSM) at temperature of 2\,K.  
		For thermal and magnetic measurements we used a sample with mass of 6.66 mg.
			
		In order to implement the Andreev reflection (AR)  spectroscopy, the SnS contacts were prepared in situ using  the break-junction technique \cite{Ponomarev1995, Kuzmicheva2014}. 
		In this technique,  a flat thin crystal is mounted on a flexible substrate and then is finely bent until it cracks at the preliminary notched narrowing. 
		As a result, a cleft at the designated location  is formed in clean cryogenic environment. 
		After cleaving, using a fine micrometric screw, we adjusted the cryogenic cleavage until a desired Andreev- or Josephson-type $I-V$ characteristic was obtained.  
		During the cryogenic experiment we tuned mechanically the cleavage, and probed, respectively, different micro-contacts in order to select a  high-transparency  contact with the required type $I-V$ characteristic. 
		In order to take four-contact  transport measurements the sample was placed on top of the thin printed circuit board (PCB) made of FR-4 textolite; the PCB, in turn, was fixed on a flat Be-bronze  springy holder, made of  0.1\,mm thick foil. 
		The sample was fixed to PCB with In-clamping contacts, reinforced with silver paste. 
		The springy holder was finely bent by a  tip of the micrometric screw  pushing the holder  from the side  opposite to the sample. 
		This design enabled us to cleave the sample and precisely vary the  distance between the cleft banks during the cryogenic experiment.
			
		It is a common knowledge, that the $dI/dV$ characteristics of  high transparency superconductor - normal metal - superconductor (SnS) microcontacts exhibit a so called subharmonic gap structure \cite{Octavio1978}, comprising characteristic dips at bias voltages $V_n= 2\Delta/ne$ with $n$ being the subharmonic number. 
		The subharmonic order $n$  equals to the number of Andreev reflections the carriers experience when moving in the normal constriction of the SnS junction. 
		Therefore, the superconducting gap value $\Delta$ may be determined directly from the measured $V_n$ versus $1/n$ dependence.
			
		The $dI/dV$ measurements were made in the temperature range from 1.5 to 4\,K. 
		The measurements  were taken by four probe technique. 
		In order to bias the sample we used a voltage controlled current source fed with a sum  of the DC bias and AC modulation voltages. 
		The running current and voltage  were measured independently using a precise digitizer; $dI$, and $dV$ were obtained using two lock-in amplifiers SR-830.

	\section{\label{sec:Results}Results and discussion}

		\subsection{\label{sec:DFTGGA}	DFT/GGA calculations}

			\begin{figure*}[t]
				\center{\includegraphics[width=14cm]{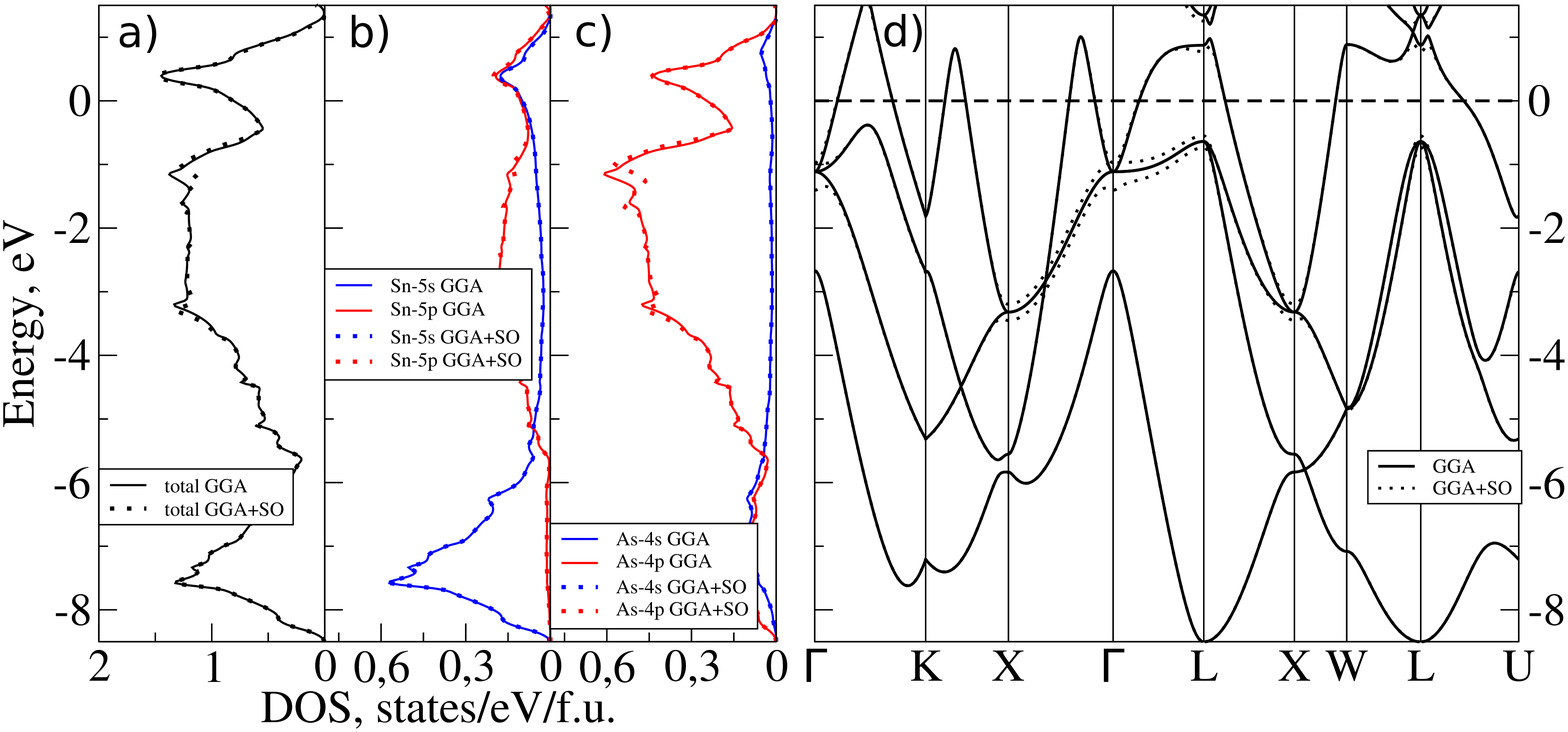}}
				\center{\includegraphics[width=14cm]{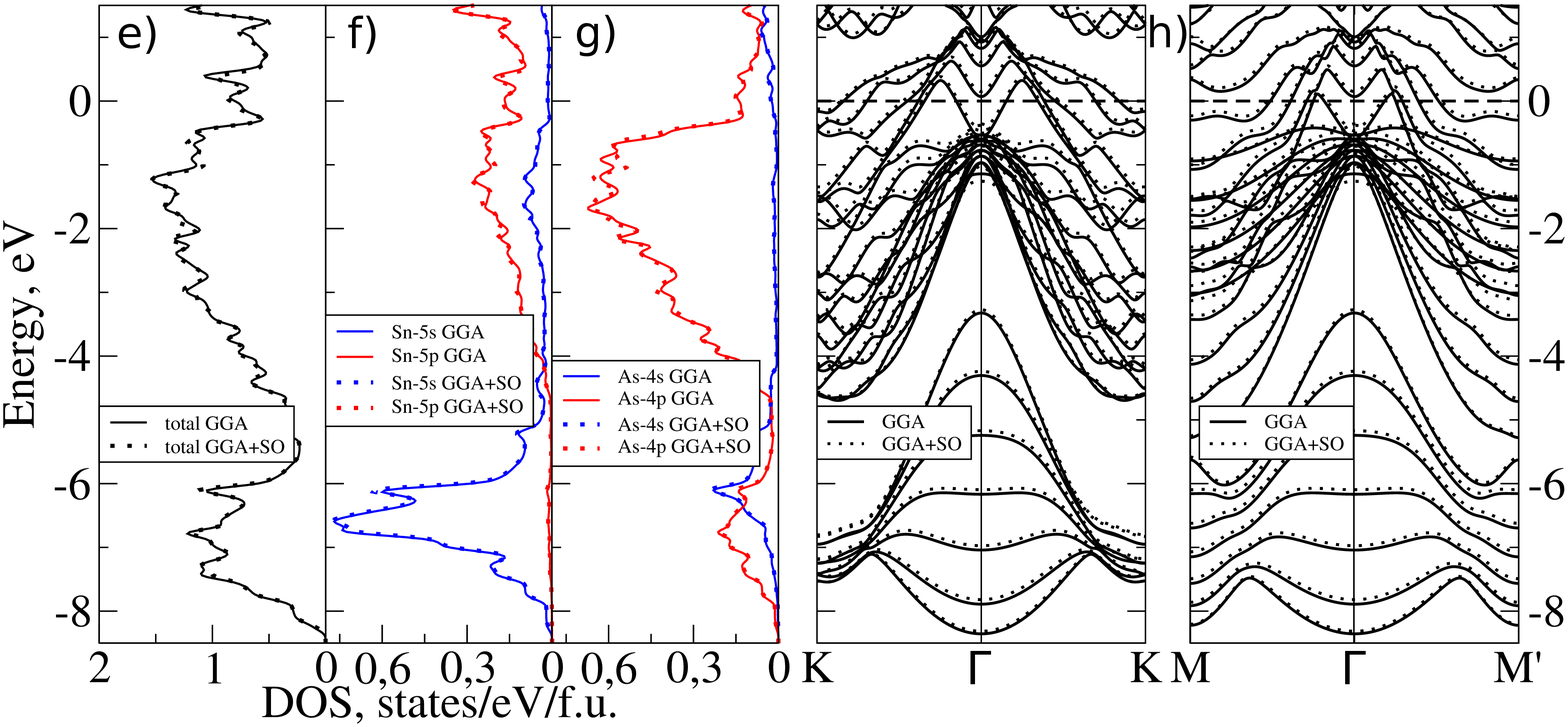}}
				\caption{DFT/GGA calculated total and partial densities of states [panels (a),b), c), e), f), g)] and band  dipersions for the bulk [panel d)] and for the slab [panel h)] crystal structures. Dotted lines denote the DFT/GGA calculations with SOC and solid lines -- without SOC. Zero energy corresponds to the Fermi level. }
				\label{fig:bands_dos}
			\end{figure*}
			
			Theoretical band structure for SnAs was calculated within density functional theory (DFT) with the full-potential linear augmented plane-wave framework as implemented in WIEN2k~\cite{wien2k} together with the generalized gradient approximation (GGA) by Perdew, Burke and Ernzerhof~\cite{pbe} to the exchange-correlation functional. 
			These calculations were converged self-consistently on a grid of 1000 \textit{k}-points in the irreducible Brillouin zone. 
			Calculations were performed both, with and without spin-orbit coupling (SOC). 
			The SOC was considered in a second-variational procedure \cite{Singh2006}. 
			Presented below DFT/GGA results for bulk SnAs agree well with those reported earlier in Refs. \cite{Tutuncu2015,HASE2016}
			
			The DFT calculations for bulk SnAs were based on the space symmetry group $Fm\overline{3}m$ (No. 225) and lattice constant obtained in this work.  
			Also, to consider real experimental geometry of the samples we have constructed (111) slab  with topmost layer of Sn (vacuum gap between slabs is 25 \AA). 
			The (111) slab crystal structure is found to have monoclinic $C2/m$ (No. 12) space symmetry group. 
			Calculations were performed for slabs with thickness of 5 and 2 unit cells (UC). 
			No difference between the band structure was found for
			5UC and 2UC slabs. 
			Correspondingly, all discussed below results were obtained for 2UC (111) slab. 
			Also, we have done structural relaxation for 2UC SnAs (111) slab, but have not seen any significant atomic position changes. 
			The relaxation causes only minor modification of the electronic structure and is not exploited furthermore.
		
			The DFT/GGA results for SnAs are presented on Fig. \ref{fig:bands_dos}.  
			Panel a) shows the total, and panels  b) and c) - partial  density of states for bulk SnAs. 
			A mixture of various electronic states (Sn-5s, Sn-5p and As-4p) is observed near Fermi level with a slight predominance of As-4p states. 
			As a result, the  Fermi level lies in the recess of density of states. 
			
			SOC impact on the DOS is most pronounced for As-4p states in the interval between -0.5 eV and -2 eV as  follows from the dotted line on panels c) and g). 
			Also, SOC slightly affects the Sn-5p states. 
			The same effect is visible for dispersions along X-$\Gamma$-L directions on panels d) and h) of Fig.  \ref{fig:bands_dos}. 
			Other bands are practically not affected by SOC. The strongest manifestation of SOC can be observed near $\Gamma$-point at about -1 eV and provides lifting of the bands degeneracy with the splitting by a  few tens of eV, in a qualitative agreement with earlier calculations   \cite{HASE2016, Bezotosnyi2017}. 
			Concerning possible link between SOC and superconductivity in SnAs, one can see that SOC doesn't manifest considerably in the vicinity of the Fermi level. 
			Therefore, one  should not expect strong influence of SOC on superconductivity.
			
\begin{figure}[h]
				\center{\includegraphics[width=0.8\linewidth]{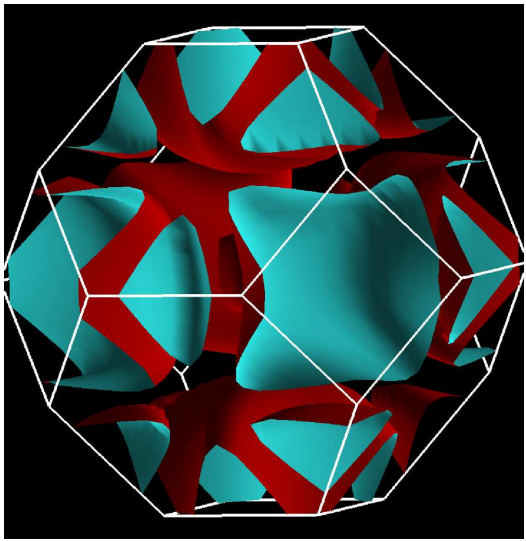}}
				\caption{SnAs DFT/GGA calculated Fermi surface.}
				\label{fig:fermisurf}
			\end{figure}

			Several observations based on the  DFT/GGA results should be mentioned regarding the  superconductivity character in SnAs. 
			Firstly, bulk DFT/GGA band structure of SnAs (Fig. \ref{fig:bands_dos} (d)) has many bands crossing the Fermi level. 
			It supposes SnAs to be a multiband superconductor and, in the normal state,  a good three-dimensional metal. 
			The shape of the calculated Fermi surface presented on Fig. \ref{fig:fermisurf} confirms the last two statements. 
			Thirdly, the total density of states at the Fermi level is rather small -- 0.81 states/eV/f.u. (f.u. - formula unit) which might lead to experimentally observed low\,$T_c$.

			From Fig. \ref{fig:bands_dos} (d) and Fig. \ref{fig:fermisurf} one can see that electronic spectra have a complicated anisotropic character. 
			There are 10 different Fermi level crossings, each has its particular Fermi momentum $k_F$ and effective electron mass $m^*$. 
			Since SnAs with a large Fermi surface is a pretty good metal, for furher analysis we employ the Fermi liquid type description and deduce from DFT/GGA calculations several parameters important for the superconducting state. 
			Firstly, we estimated the effective  mass enhancement  through the specific heat linear coefficient $\gamma_{calc}=\frac{\pi^2}{3}N_A k^2_B N(E_F)$=1.89 mJ/(mol$\times$K$^2$), where $N_A$ is the Avogadro constant, $k_B$ -- Boltzmann constant and $N(E_F)$ -- the DFT/GGA value of total DOS at the Fermi level. 
			Then  $\frac{\gamma_{n}}{\gamma_{calc}}=\frac{m^*}{m_e}$ with $\gamma_{n}$=2.67 mJ/(mol$\times$K$^2$) -- electronic specific heat in the normal state (see below) gives $\frac{m^*}{m_e}$=1.41. 
			This ratio agrees well with Ref. \cite{HASE2016} despite both calculated $\gamma_{\rm calc}$ and experimental $\gamma_n$ exceed by a factor of $\sim$ 1.2 the respective values used in  Ref. \cite{HASE2016}.
			
			Specific heat,  effective mass and  Fermi momentum  may be characterized,  to the first approximation, with their single values averaged over the Fermi surface despite complicated multi band character of electron dispersions. 
			Finding such averaged parameters  for arbitrary non-elliptic Fermi surface  (see Fig. \ref{fig:fermisurf}) is rather difficult. 
			One can try to obtain $k_F$ value again using $N(E_F)$ -- the  calculated value of the total  DOS at the Fermi level. 
			Then $k_F = N(E_F)\frac{4 \pi^2 \hbar^2}{V \frac{m^*}{m_e}m_e}=0.6$\AA$^{-1}$, where $V$ is elementary cubic cell volume. 
			Also it may be found as $k_F=(3\pi^2n)^{1/3}$, where $n=N/V$ with $N$ -- number of valence electrons in the unit cell. 
			From partial DOS one can see that there are 2.1 of As-4p electrons, 0.1 Sn-5s electrons and 0.72 Sn-5p electrons; hence, in total  $N$ per formula unit is 2.92. 
			There are 4 formula units per cubic cell with $V$=187.5\,\AA$^3$. 
			It gives $k_F$=1.23\,\AA$^{-1}$. 
			Such discrepancy between $k_F$ values comes from multiband anisotropic nature of SnAs spectra while all definitions above are for the isotropic single band Fermi liquid. 
			There is also theoretical conjecture for concentration in Ref. \cite{Geller1964} n=2.14$\times$10$^{22}$\,cm$^{-3}$, which gives $k_F$=0.86\,\AA$^{-1}$.
			
			Previously these values were estimated in the Ref.~\cite{Wang2014} to be $m^*=1.20m_e$ and $k_F=1.24$\,\AA$^{-1}$ respectively within a straightforward single-sheet spherical Fermi surface approach. 
			Here, from the DFT/GGA calculations we see that the Fermi surface has many sheets and is rather anisotropic. 
			Although the values of  $m^*$ and $k_F$ obtained here on the basis of DFT/GGA calculations are surprisingly quite similar to those of the Ref.~\cite{Wang2014}.
		
			Lower panels e) to h) in Fig. \ref{fig:bands_dos} present DFT/GGA results for (111) slab. 
			One can see that the total and partial DOS are practically the same as those for bulk SnAs. 
			The value of total DOS at the Fermi level also is nearly the same as the bulk one. Manifestations of SOC are weak, however the   band dispersions are rather different. 
			First of all, the difference comes from different paths in $k$-space for bulk and (111) slab as shown on Fig.~\ref{fig:ARPES}(a). 
			Secondly, there are much more bands for the slab since $k_z$ translation invariance is broken for the surface. 
			As a result, all atoms even of the same sort become inequivalent and produce their own set of bands. 
			Nevertheless, in general SnAs bulk band shape to some extent resembles that for (111) slab crystal structure.

		\subsection{\label{sec:ARPES}	ARPES vs. DFT/GGA}
				
			Earlier ARPES results obtained by our group were reported in Ref.~\cite{Bezotosnyi2017}. 
			We noted there the observed  band bundle spliting at the $\Gamma$-point not captured  by  the existed DFT calculations \cite{Tutuncu2015, HASE2016}. 
			We suggested  this disagreement to be caused either by SOC, or by a  band structure 	reconstruction on the crystal surface \cite{Bezotosnyi2017}. 
			We now compare ARPES data and DFT/GGA calculated band dispersions to clarify these issues and conclude  that the reason of the band splitting is indeed the reconstruction of the band structure at the crystal surface \cite{Bezotosnyi2017}.
				
			\begin{figure*}[t!]
				\includegraphics[width=16cm]{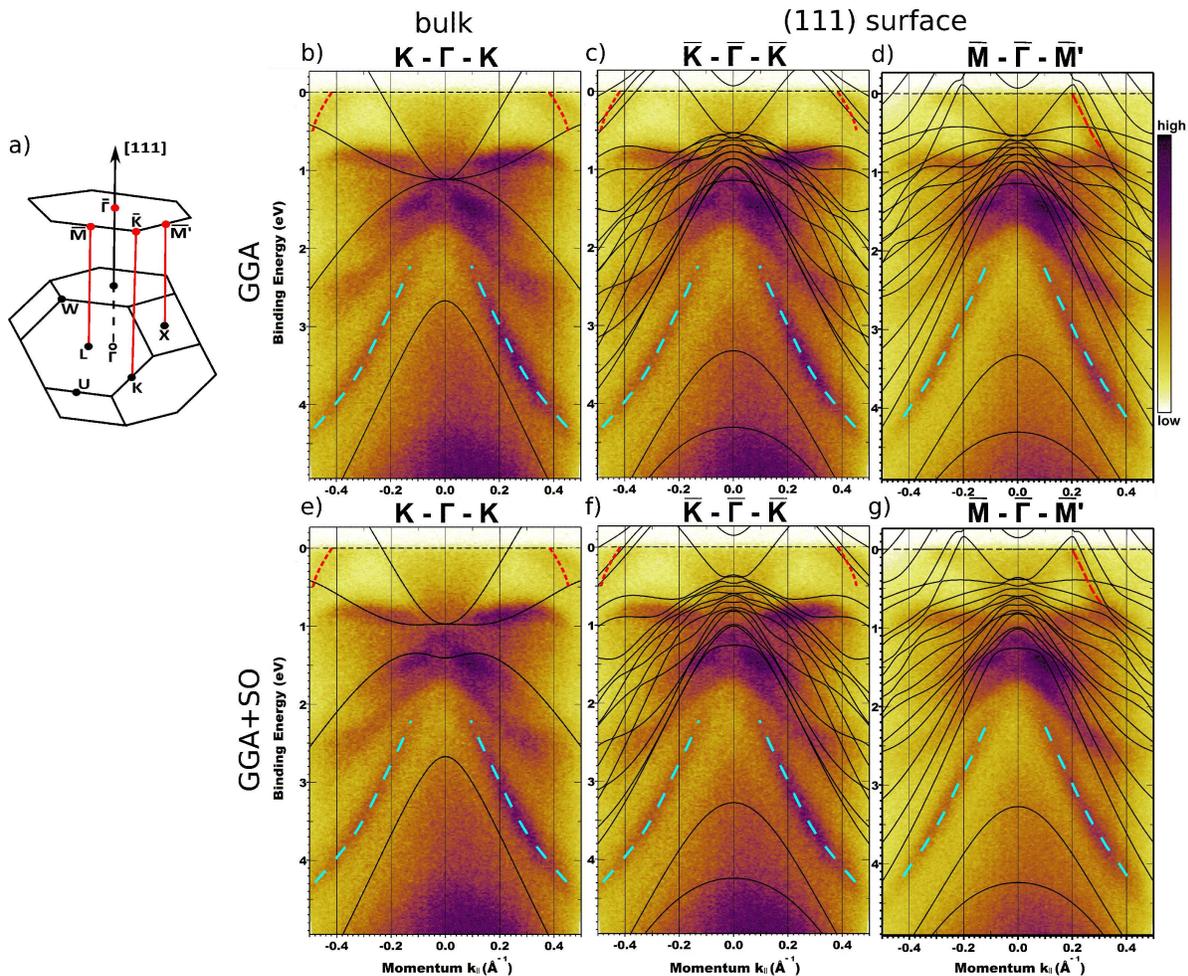}	
				\caption{a) The first Brillouin zone for the 	NaCl-type structure with  the projection to the (111) plane.  b), e) calculated bulk band structure (solid lines) and ARPES spectra (contour plots) for SnAs along $\mathrm{K-\Gamma-K}$ direction. c), d), f), g) ARPES spectra with DFT/GGA calculated band structure of (111) slab along high symmetry directions  $\mathrm{\bar{K}-\bar{\Gamma}-\bar{K}}$ and $\mathrm{\Bar{M}-\Bar{\Gamma}-\Bar{M'}}$ from panel (a). The lower row corresponds to DFT/GGA calculations with SOC. Dashed lines near Fermi level show maxima of the ARPES data maps. Zero energy corresponds to the Fermi level.}
				\label{fig:ARPES}
			\end{figure*}
		
			Bulk calculations for $k_z$ other than 0 (the $k_z$ direction coincides with the [111] direction) show, that by varying $k_z$ it is not possible to improve  agreement between the calculated and experimental spectra, similar to a number of studies \cite{Uruma2007, Damascelli2003}. 
			In this regard, the work presents the calculation results for the case $k_z$ = 0.
		
			In Fig. ~\ref{fig:ARPES} panels (b) and (e) show ARPES data and DFT/GGA bands for bulk SnAs without and with SOC, respectively. 
			The SnAs bulk DFT/GGA bands resemble ARPES data very remotely. 
			There is a bundle of bands around 1.1 eV (predominantly of As-4p character) in calculated and ARPES data. 
			However, DFT/GGA shows only three bands in the bundle plus one band by 1.5eV lower, whereas  ARPES clearly demonstrate four bands in the vicinity of the $\Gamma$-point. 
			Also energy positions of those bands are quite different in ARPES and DFT/GGA. 
			Somehow ARPES does not resolve bands crossing the Fermi level, the most intensive signal is in the vicinity of the $\Gamma$-point (the momenta interval from -0.2 to 0.2 \AA$^{-1}$ of the ARPES spectra, Fig.\,\ref{fig:ARPES}). 
		
			Inclusion of SOC splitting slightly improves the situation. 
			The wings around 0.9 and 1.3 eV agrees a bit better with experiment but ARPES does not show clear separation of those bands. 
			One should note also that bulk DFT/GGA bands  along $\mathrm{K-\Gamma-K}$ direction do not reproduce ARPES bands near Fermi level shown by red dashed curves on ARPES spectra.
		
			In case of (111) slab calculations, as panels (c,d,f,g) of Fig.\,\ref{fig:ARPES}   show, the calculations demonstrate  the emergence of electronic dispersion which much better reproduce ARPES bands near the Fermi level shown with red  dashed lines for both  $\mathrm{\bar{K}-\bar{\Gamma}-\bar{K}}$ and $\mathrm{\Bar{M}-\Bar{\Gamma}-\Bar{M'}}$ directions.  
			We conclude, the Fermi level crossing positions are well captured by DFT/GGA calculations in the vicinity of the Fermi level. 
			Wings at about 0.9 eV are well reproduced here too. Dark ARPES data region at $\Gamma$-point about 1.1 eV below $E_F$ is qualitatively represented by bunch of DFT/GGA ``spaghetti''. 
			Strictly speaking not all of these ``spaghetti'' may be seen by ARPES since at a given incident beam energy the  penetration depth is about 7\AA. 
			It corresponds to one, maximum two top most surface layers. 
			The layers below should not provide significant ARPES signal. 
			SOC has almost no effect on the shape of the DFT/GGA bands and does not bring about agreement with ARPES data. 
			Based on the above  comparison we  conclude, that the ARPES spectra in general in the vicinity of $E_F$ are in good agreement with the DFT/GGA calculated bands for (111) slab of SnAs. 
			However, the only one band with maximum at about 3\,eV (highlighted with light blue dashed curves in Fig.\,\ref{fig:ARPES}) isn't reproduced by any of the performed calculations. 
			We suppose, that this feature might be a main theme of further studies.

		\subsection{\label{sec:SC}Resistivity, specific heat and magnetic susceptibility}

\begin{figure}[h]
				\includegraphics[width=8cm]{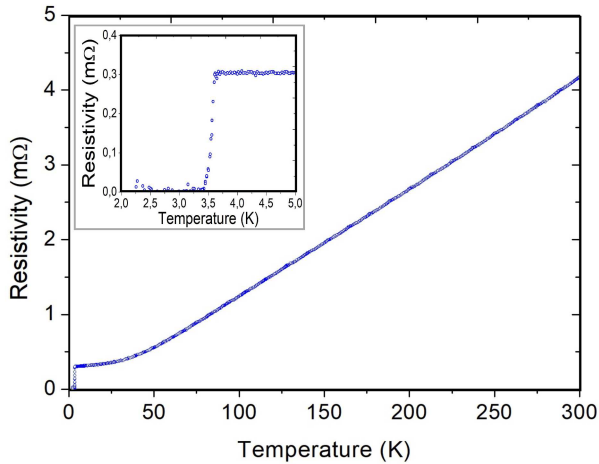} 	
				\caption{Temperature dependence of resistivity for SnAs (sample No.\,1). The inset shows the zoomed-in region of superconducting transition.}
				\label{fig:Res}
			\end{figure}

			For more detailed information about superconducting properties of SnAs we performed resistivity, specific heat and  magnetic susceptibility  measurements. 
			Temperature dependence of resistivity is shown in Fig.~\ref{fig:Res} and demonstrates the metallic-type conduction of the sample, which is consistent with the band structure calculations and ARPES data (Fig.~\ref{fig:ARPES}). 
			At a temperature of about 3.6\,K (for sample No.\,1), the resistivity shows a transition to the superconducting state (Fig.~\ref{fig:Res}, inset).  
			This agrees  with the  earlier data \cite{Wang2014}.

\begin{figure}[h]
				\includegraphics[width=8cm]{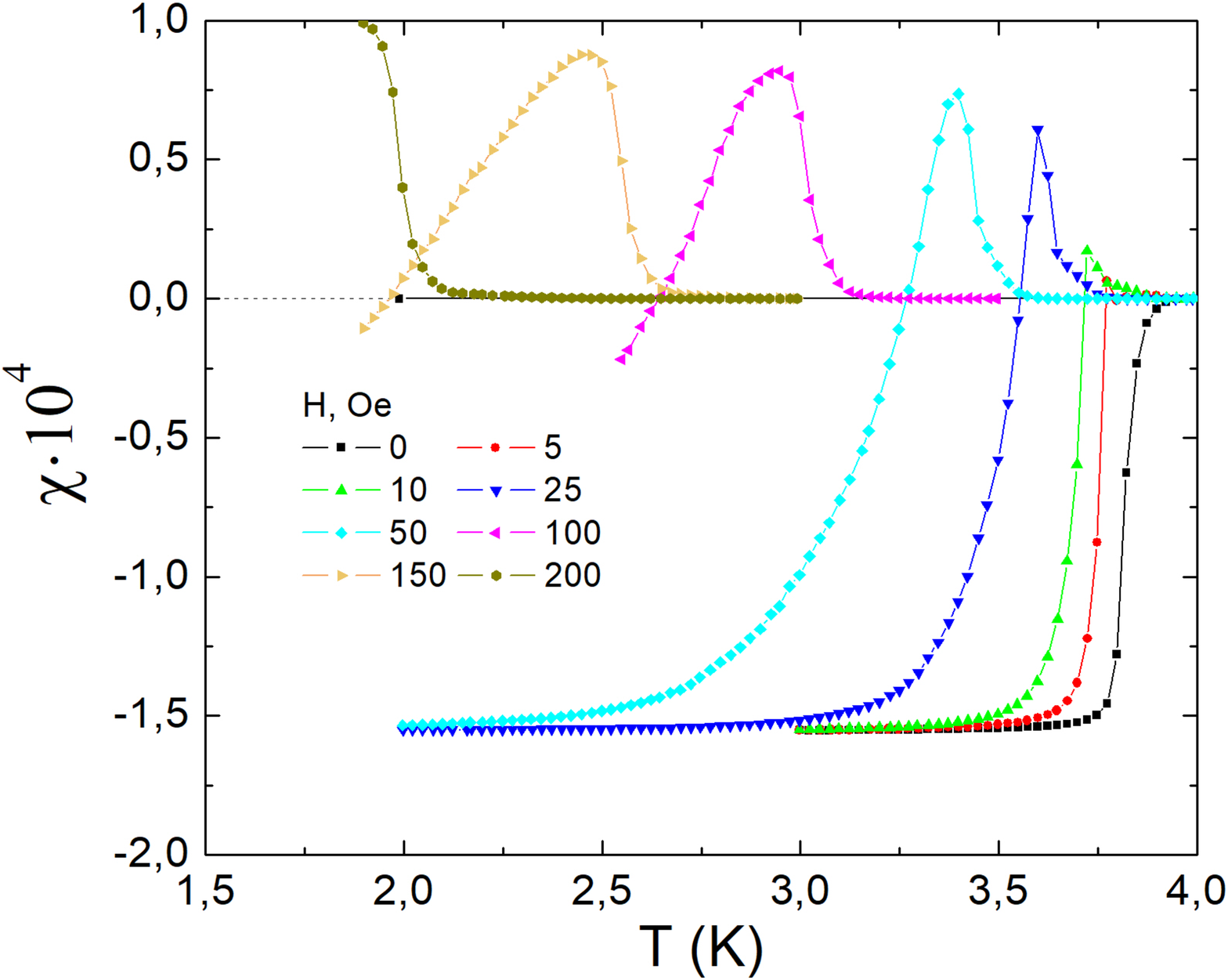} 	
				\caption{AC--Magnetic susceptibility of the SnAs (sample No.\,2).}
				\label{fig:Hi}
			\end{figure}

			Temperature dependence of the magnetic AC susceptibility  $\chi$  of SnAs crystal is presented in Fig.~\ref{fig:Hi} (for sample No.\,2).   
			The susceptibility for this sample  demonstrates superconducting (SC) transition near 3.8\,K in zero field, slightly higher than the SC transition in transport (for sample No.\,1). 
			As  the applied DC magnetic field increases, the superconducting transition temperature  gradually decreases and an additional positive peak appears in the $\chi (T)$ dependence near $T_c$. 
			This peak corresponds to the so called differential paramagnetic effect (DPE) in superconductors \cite{Hein1961, Zhao2012, Yonezawa2005}.  	
			DPE (positive $\chi_{ac}$) corresponds to the existence of a region with strong growth in $M(H)$ dependence  (and, as a consequence, positive derivative  $\partial$M/$\partial$H). 
			This effect is observed for mixed state of type-I superconductors  or in type-II superconductors with close ${H_c}_1$ and ${H_c}_2$ values. 
			
\begin{figure}[h]
				\includegraphics[width=8cm]{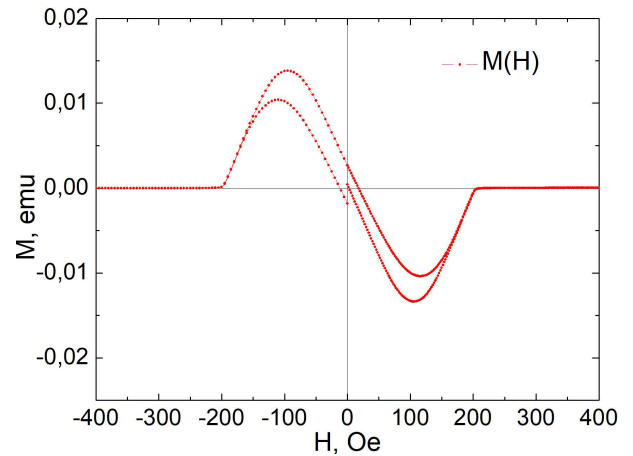} 	
				\caption{Isothermal magnetization loop $M(H)$ at $2$\,K (sample No.\,2).}
				\label{fig:MH}
			\end{figure}

			Figure~\ref{fig:MH} shows isothermal magnetization loop $M(H)$ at $T=2$\,K. 
			The shape of the curve is nearly typical for   type-I superconductor, however it reveals a small hysteresis. 
			Such hysteresis is observed in other compounds considered to be type-I superconductors and, generally,  is associated with sample shape effects, domain or grain walls etc \cite{Zhao2012, Sun2016, Yamaguchi1987, Anand2011, Tran2013}. 
			
\begin{figure}[h]
				\includegraphics[width=8cm]{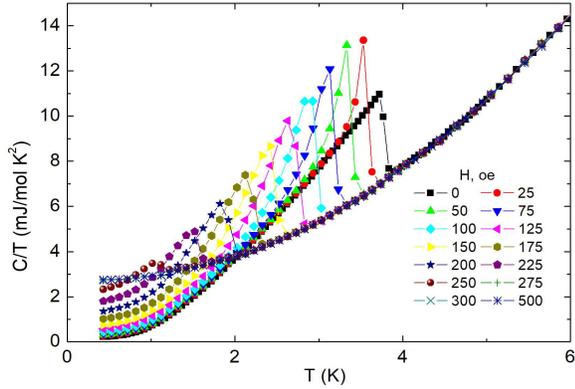} 	
				\caption{Temperature dependences $C(T)/T$  for the SnAs crystal (sample No.\,2) in magnetic fields 0 -- 500 Oe.}
				\label{fig:SHt}
			\end{figure}

			Temperature dependences of the  specific heat   in various magnetic fields are shown on Fig.~\ref{fig:SHt}. 
			The sharp jump in specific heat at $H=0$ for temperature near 4\,K confirms the bulk superconducting transition.  
			From the local entropy conservation  we find the critical temperature $T_c = 3.80 \pm 0.05$\,K,  consistent with susceptibility measurements. 
			The total width of the  SC transition $\Delta T_c = 0.15$\,K is another evidence of the high quality of the studied crystal. 
			In general, our data for zero field specific heat is reasonably consistent with   \cite{Wang2014}. 

			Specific heat tends to zero at $T\longrightarrow 0$ in zero field and does not demonstrate any anomalies in the low-temperature region. 
			This tendency is an additional evidence of bulk nature of superconductivity in the studied samples. 
			The polynomial extrapolation to zero temperature $C(T)/T=\beta T^2+\gamma_r$ ($\beta$=0,317$\pm$0,019 mJ/(mol$\times$K$^4$)) gives residual electronic specific heat $\gamma_r=0.015$\,mJ/(mol$\times$K$^2$).

			Magnetic field suppresses the superconducting transition and shifts it gradually to lower temperatures. 
			At fields above 300\,Oe, the superconductivity is fully suppressed, as follows from the equality of the specific heat at $H = 300$\,Oe and higher fields, for example, 500\,Oe. 
			Thus, the electronic component of the specific heat can be defined as $C_e = C(H) - C(H_0)  +\gamma_nT$  for $H<H_0$, (where $H_0=300$\,Oe exceeds the critical magnetic field), taking into account the entropy conservation.  
			The electronic specific heat in the normal state $\gamma_n$ at temperatures above $T_c$ was found to be $\gamma_n= 2.67$\,mJ/(mol$\times$K$^2$).
			
			\begin{figure}[h]
				\includegraphics[width=8cm]{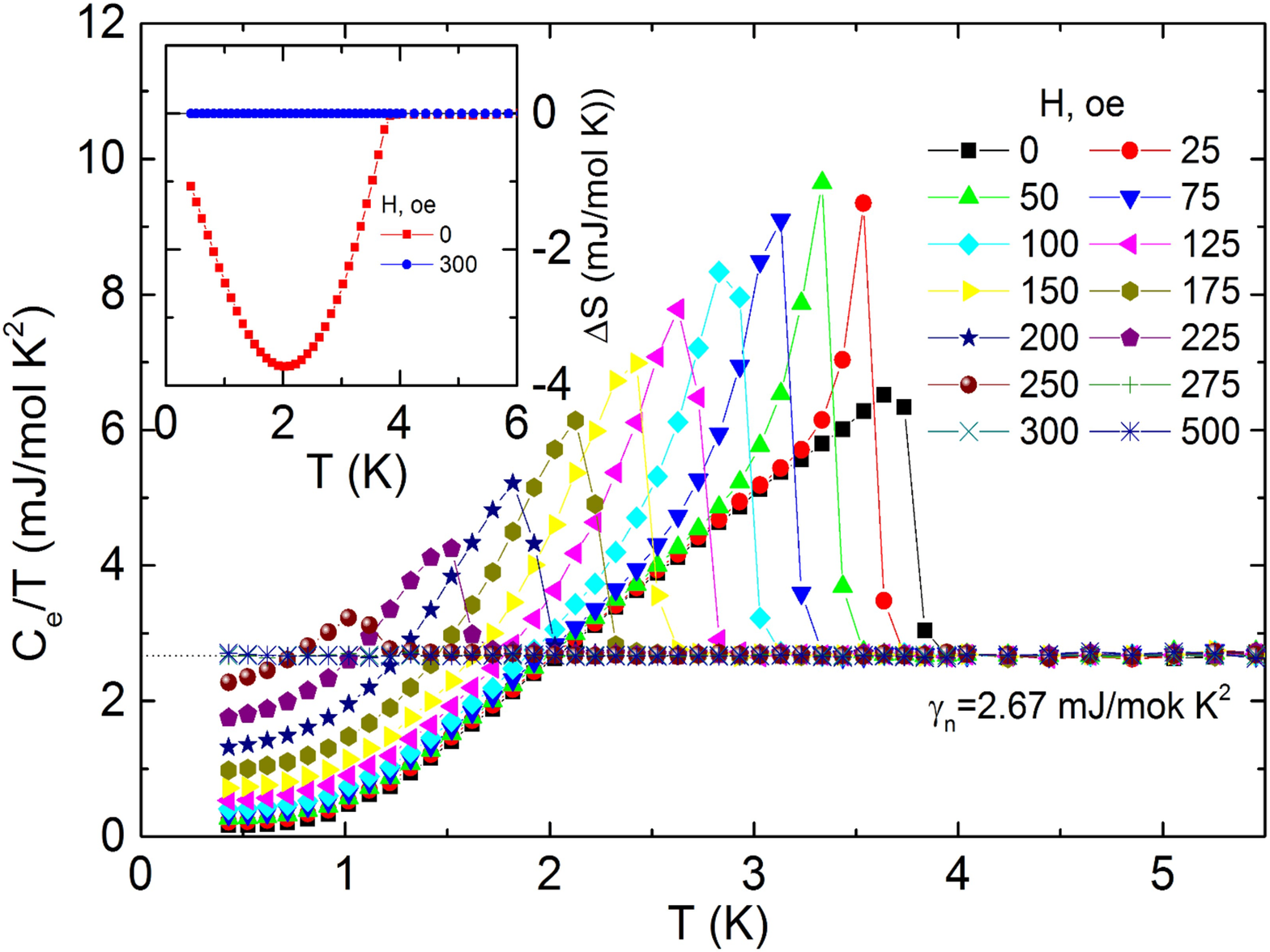} 	
				\caption{Electronic specific heat $C_e(T)/T $ in fields 0 -- 500\,Oe (sample No.\,2). {\itshape{Inset}}: the difference between the entropy in superconducting state at zero field and normal state (at $H=300$\,Oe) $\Delta S = S_s-S_n$.}
				\label{fig:CeT}
			\end{figure}
				
			The residual term $\gamma_r$ is much less than the electronic specific heat in the normal state $\gamma_n$. 
			The ratio $\gamma_r/\gamma_n \approx 0.5\%$ quantifies the relative amount of  unpaired carriers; it confirms the bulk character of the superconducting state, and evidences for a high crystal purity.

			Temperature dependence of the electronic contribution to specific heat $C_e(T)/T $ in various fields is shown in Fig.~\ref{fig:CeT}.
		
			Inset of Fig.~\ref{fig:CeT} shows difference between entropy in superconducting and normal states. 
			We now estimate critical magnetic field value from the obtained data. 
			It could be done from equation $w=\frac{|\Delta F|\cdot N}{V}=\frac{H_c^2}{8\pi}$, where $w$ is the density of energy, $\Delta$F is the difference between free energies in superconducting and normal states, N is the molar quantity, and V is the sample volume. 
			Numerical integration give us $\Delta F = \int_{0}^{T_c} \Delta SdT$= -- 9.00 mJ/mol, where $\Delta S$ is the difference between the entropy in superconducting state at zero field and in normal state. 
			Taking into account N=m/$\mu$=3.44$\cdot$10$^{-5}$ mol (here $\mu$ is the formula mass of SnAs, m= 6.66 g is the sample mass) V =m/$\rho$= 0.971$\cdot$10$^{-3}$ cm$^3$ ($\rho$=6.86 g/cm$^3$ is the mass density) we have for critical magnetic field H$_c$=283 Oe.
		
			\begin{figure}[]
				\includegraphics[width=8 cm]{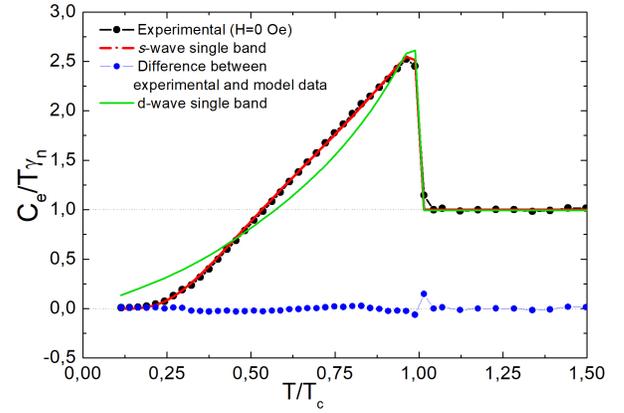}	
				\caption{ Normalized electronic specific heat of the superconducting condensate $\dfrac{C_{e}}{T \gamma_n }$ and its best fit with   the s- and d- wave $\alpha$-model (sample No.\,2). }
				\label{fig:CTn}
			\end{figure}

			To analyze thermal properties of superconducting condensate, it is convenient to consider the normalized electronic specific heat $C_{en} = \dfrac{C_e}{T\gamma_n}$. 
			For the superconducting condensate it may be calculated within the framework of the BCS theory using the so-called alpha model \cite{Bouquet2001}. 
			The $\alpha$-model may be generalized in case of possible in-plane anisotropy (d-wave $\alpha$-model) as presented in \cite{Muratov2018},  with two adjustable parameters $\alpha =\dfrac{2\Delta(0)}{k_BT_c}$, ($\Delta(0)$ is the superconducting gap at zero temperature), and  $m(\varphi)=1 + \mu \cos (2\varphi)$ -- angular dependence of the gap.

			Normalized electronic specific heat of the superconducting condensate $\dfrac{C_{e}}{T \gamma_n }$ versus $(T/T_c)$ and its best fit with   the s- and d- wave $\alpha$-model \cite{Bouquet2001} are presented in Fig.~\ref{fig:CTn}.	
			The simplest s-wave model (isotropic gap, $m (\varphi) = 1$) has only one free parameter $\alpha$. 
			The best description of the experimental data was obtained with $\alpha$ = 3.73 which is very close to the characteristic BCS ratio ($\alpha$ = 3.52) and corresponds to the superconducting gap of $\Delta=0.6$\,meV (see Fig.~\ref{fig:CTn}). 	
			Clearly, the model with the d-wave  gap symmetry ($m(\varphi) = \cos 2\varphi$)  is at odd with the experimental data  of Fig.~\ref{fig:CTn}.  
			Fitting with the  extended s-wave gap symmetry ($m(\varphi)=1 + \mu \cos (2\varphi)$)  leads to $\mu \rightarrow 0$, and therefore reduces to the simple s-wave model with an isotropic gap. 
			The same result is obtained for the two-band $\alpha$-model which corresponds to the two-band superconductivity \cite{Bouquet2001}:	$C(T)=\varphi_1C_1(T)+\varphi_2C_2(T)$. 
			In this case from the fitting procedure, the contribution of one of the gaps $\varphi_1$ becomes equal to zero, whereas angular modulation for the second one vanishes ($\mu\longrightarrow 1$).

			Evidently, the best description of the experimental data is obtained using the single-band BCS model with an isotropic gap and $\alpha=\dfrac{2\Delta(0)}{k_BT_c}=3.73$. 
			Despite SnAs as shown above has essentially multi band electronic system application of single band BCS model can be justified in case matrix of coupling constant consists of approximately equal values for all Fermi surface sheets \cite{KUCHINSKII2010}.

			Figures ~\ref{fig:SHt} and \ref{fig:CeT}	show temperature dependence of specific heat in magnetic field. 
			Suppression of the specific heat jump  at H=500\,Oe was also reported in Ref. ~\cite{Wang2014},	 however, the  temperature dependence of specific heat in lower magnetic fields   wasn't  explored there. 
			For nonzero  fields, beside the conventional jump at $T_c(H)$,  the electronic specific heat demonstrates an additional sharp peak near the SC transition (see Figs.~\ref{fig:SHt} and \ref{fig:CeT}). 
			Commonly, such feature is considered as a transformation of the 2nd- to the 1st- order (in the presence of a magnetic field) phase transition in type-I superconductors, such as, e.g., thallium and  aluminum  \cite{Misener1940, Rorer1963} as well as  for type-I compound superconductors  \cite{Sun2016, Zhao2012, Anand2011}.  
			The increase in the specific heat near the superconducting transition signifies, that an additional energy is required to realize the superconducting transition in magnetic field.  

			Similar behavior  may be  also observed for  other first-order phase transitions:  e.g., for melting transition,  where the specific heat of a crystal changes to the specific heat of a liquid right at the melting point, however, the temperature does not change during melting despite the heat enters the system; as a result,  the specific heat  exhibits a sharp peak. 
			In our case, similarly, the  transition from superconducting to normal state occurs with the absorption of latent heat. 
			The finite width of the peak is due to the intermediate state (shape effect). 
			Thus, the specific heat   in magnetic fields demonstrates features intrinsic to the first-order phase transition and indicates SnAs to be the type-I superconductor. 
	
			\begin{figure}[]
				\includegraphics[width=8 cm]{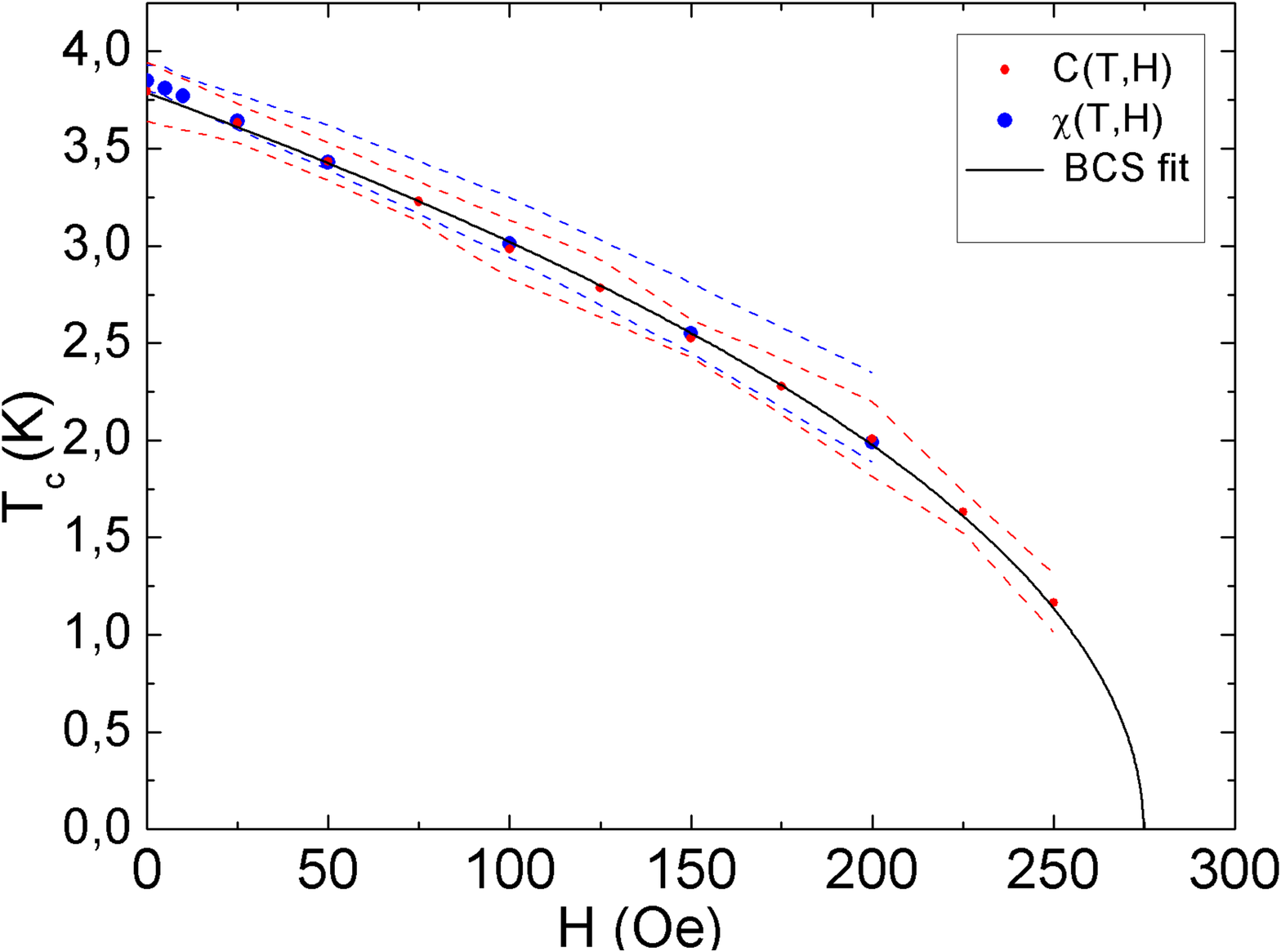}	
				\caption{ Critical temperature $T_c$ vs magnetic field, determined from  the AC susceptibility and specific heat; solid line represents  the BCS parabolic fit $H_c(T) = H_c(0)(1 - {(T / T_c)}^2)$ (sample No.\,2).}
				\label{fig:PT}
			\end{figure}

			Figure~\ref{fig:PT} shows magnetic field dependence of the superconducting transition temperature, determined from  the AC susceptibility and specific heat data. 
			The dashed lines  depict the corridor between the beginning and ending of the transition (red lines  and dots correspond to the specific heat data and blue ones -- to the AC susceptibility). 
			The points correspond to the mean values. 
			The mean $T_c(H)$ data obtained from the specific heat coincides with that obtained from the susceptibility measurements  and slightly exceeds the resistivity data. 
			This dependence is well described by the BCS dependence $H_c(T) = H_c(0)(1 - (T / T_c)^2)$ that extrapolates to  $H_c(0) = 275$\,Oe; the latter value is reasonably consistent with $H_c=283$\,Oe, estimated from the difference in entropy between superconducting and normal states.
		
			The data obtained  enables one to estimate superconducting parameters of the SnAs compound. 
			Electronic part of specific heat in normal state  together with k$_F$ and $m^*$ found from DFT/GGA results described above give possibility  to estimate London penetration depth $\lambda_L$, coherence length $\xi$, and  Ginzburg-Landau parameter $\kappa$. 
			For calculations we take DFT/GGA based Fermi momenta k$_F$ from 0.6 to 1.23\,\AA$^{-1}$,  while $m^*$=1.41m$_e$ is rather safely obtained from DFT/GGA and specific heat.  
		
			Then London penetration depth may be estimated using formula $\lambda_L(0) = (3\pi^2 m^*/[\mu_0k_F^3e^2])^{1/2}$. It gives  $\lambda_L(0)$ in the range from 25.7 to 75.4 nm. 
			Corespondingly, coherence length $\xi(0) = 0.18 \hbar^2k_F/(k_BT_cm^*)$ equals to   171.0 -- 350.5\,nm \cite{Orlando1979} and Ginzburg--Landau parameter $\kappa = \lambda_L(0)/\xi(0)$ = 0.07 -- 0.4 $< 1/\sqrt{2}$;  the latter confirms our conclusion on the type I superconductivity for SnAs. 
			For a particular choice $k_F$=1.23\,\AA$^{-1}$, our results are close to those of Ref.~\cite{Wang2014}.

		\subsection{\label{sec:AS}Andreev reflections spectroscopy}

			In order to have a deeper insight into the superconducting properties of SnAs and, particularly,  temperature  dependence  of  the superconducting energy gap $\Delta$, we  measured  Andreev reflection spectra. 	
			The superconductor - normal metal - superconductor (SnS) Andreev reflection spectroscopy is a powerful  tool to determine the energy gap value and its temperature evolution. 
			Unlike several other intimate techniques, such as scanning tunneling spectroscopy (STS), and point-contact Andreev reflections (PCAR), this technique does not require any  data fitting \cite{Kummel1990, Klapwijk1982} or a model for data interpreting. 
			Recently, Andreev reflections spectroscopy on symmetric SnS contacts was successfully used to quantify the gap structure for  several multi-gap superconducting arsenides \cite{Kuzmicheva2017, Abdel2014, Abdel2018}.
			
			\begin{figure}[h!]
				\includegraphics[width=8 cm]{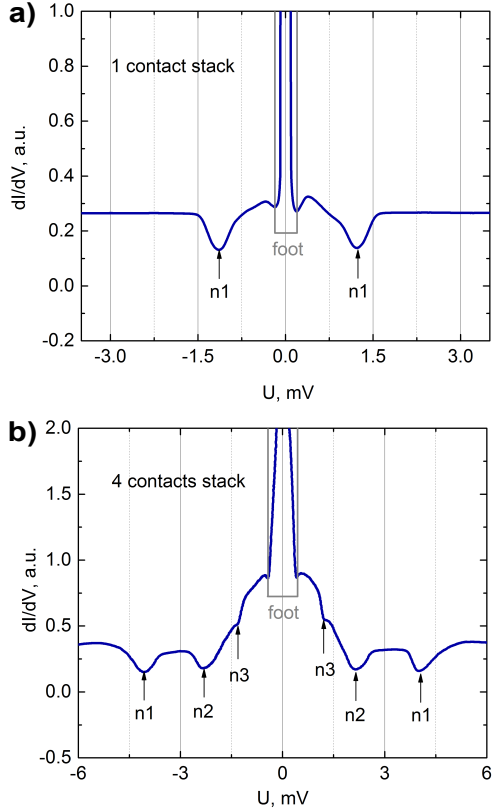}	
				\caption{ Dynamic conductances, measured at T=1.5K for  a single SnS contact  (a)  and for a 4-contact stack (b) (sample No.\,1).}
				\label{fig:AS}
			\end{figure}
			
			Figure~\ref{fig:AS}(a) shows the dynamic conductance $dI/dV$ of a single  Andreev-type contact, measured at $T=$1.5\,K. 
			The two symmetric  dips in $dI/dV$ marked with label $n_1$ in Fig.~\ref{fig:AS}(a) correspond to bias voltages $V_1=\pm 2\Delta/n_1 e$. 
			No other features are seen at higher bias, and we conclude that the dips correspond to the major  $n_1=1$ resonance. 
			From this representative data we obtain an estimate  $\Delta \approx 0.55$\,meV. Andreev reflection dips of higher order may not be seen on this contact due to the following reasons. 
			According to Ref. \cite{Klapwijk1982}, the number of possible Andreev reflections is limited by the ratio $2a/l$, where $a$ - is characteristic dimension of the constriction, $l$ - is the mean free path. 
			Another possible reason is that the second and higher-order reflection dips may be masked by the sharply growing ``foot'' (excess conductance at low biases, which is intrinsic to Andreev contacts). 

			By  mechanical tuning  the sample bending and thus, the break junctions area one can realize Andreev contacts with different $2a/l$ ratio and also stacks of sequentially connected S-n-S-n...-S contacts \cite{Ponomarev1995}. 
			For the  chain of $m$ contacts, evidently, the resonant bias voltage $V_n=m\times(2\Delta/ne)$ is increased by a factor of $m$. 
			Figure~\ref{fig:AS}(b) shows dynamic conductance  for such stack of several contacts.   
			One can see 3 features, corresponding to 3 resonances. 
			Having  the preceding single-contact spectrum, as a reference, we immediately disentangle the spectrum Fig.~\ref{fig:AS}(b) and find that it is produced by a chain of 3 sequential contacts. 
			Indeed, the observed dips at  4.1, 2,15 and 1.35\,meV may be fitted with $n_1=1, n_2=2, n_3=3$ and the consistency with the single contact spectrum is obtained for $m=3$. 
			Finally,  based on the above data   we  refine the gap value, $\Delta=0.53 \pm 0.04$\,meV  and the extrapolated value $\Delta(T\rightarrow 0)=0.54$\,meV. 
			We stress, that for all studied contacts we didn't observe signatures of the second gap.
				
			\begin{figure}[h]
				\includegraphics[width=8 cm]{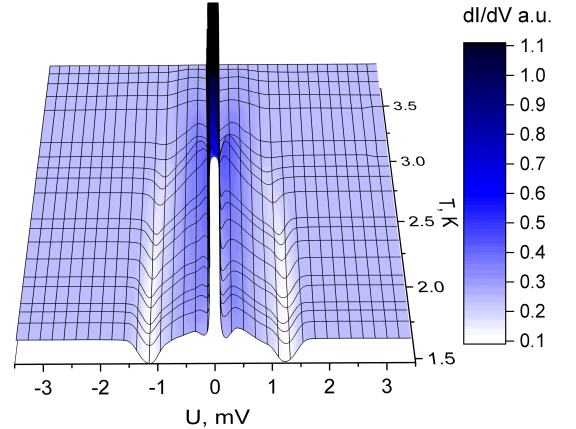}	
				\caption{ Temperature evolution of dynamic conductance for 1 contact stack a single Andreev contact (the peak at zero bias is cut off).}
				\label{fig:ASTemp}
			\end{figure}
			
			In order to find $\Delta(T)$ temperature dependence, we  measured the dynamic conductance of a single contact in the range 1.5 -- 4\,K. 
			Figure~\ref{fig:ASTemp}  shows that with temperature rising, the dips in $dI/dV$ shrink towards zero bias and the dynamic conductance finally linearizes at temperature $\approx 3.6$\,K, which is taken as the local critical temperature ${T_c}^{local}$.
		
			\begin{figure}[h]
				\includegraphics[width=8 cm]{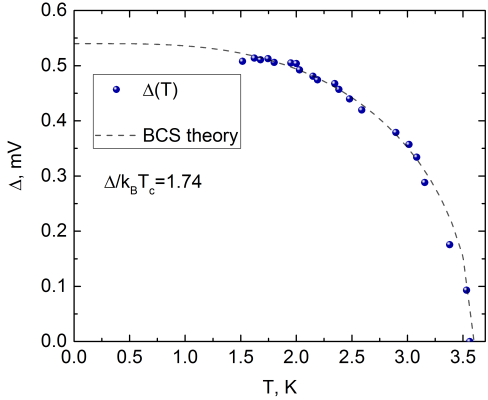}	
				\caption{  Energy gap temperature dependence deduced from the measured Andreev spectra. $\Delta=0.54 \pm 0.04$\, meV is found from extrapolation of $\Delta(T)$  to $T=0$, 	obtained with the BCS fitting dependence.}
				\label{fig:Andreev}
			\end{figure}
	
			Evolution of the gap value with temperature $\Delta(T)$ is plotted in Fig.~\ref{fig:Andreev}. 
			One can see that temperature dependence of the energy gap is well approximated by the  single-gap BCS theory,  and the ratio $2\Delta(0)/k_BT_c = 3.48$ signals the weak-coupling case.

	\section{Conclusions}
		\begin{table}[h!]
			\centering
			\begin{tabular}{l c c}
				\hline
				\hline
				Parameters & Values & Values from Ref.\cite{Wang2014}\\
				\hline
				\hline
		
				a, \AA & 5.723 & 5.72513(4) \\
		
				$\rho$, g/cm$^3$ & 6.86 & --\\
		
				$T_c$, K & 3.6 (sample No.\,1) & 3.58 \\
				   		& 3.8(sample No.\,2)& \\
				
				H$_c$, Oe & 283 (entropy) & 178 (extrapolation) \\
						& 275 (extrapolation) &  \\
		
				$\Delta(0)$, meV & 0.6 (specific heat) & --\\
								& 0.53$\pm$0.04 (AR)& \\
		
				$2\Delta(0)/T_c$ & 3.48 - 3.73 & --\\
		
				$\gamma_r$, mJ/(mol$\times$K$^2$) & 0.015 & --\\ 	
			
				$\gamma_n$, mJ/(mol$\times$K$^2$) & 2.67 & 2.18\\
		
				$\beta$,   mJ/(mol$\times$K$^4$) & 0.317 & 0.30 \\
	
				\hline
		
			\end{tabular}
			\caption{\label{tab:parameter}  Structural and thermodynamic data for SnAs in the normal and superconducting  states.	 Parameters H$_c$, $\gamma_r$, $\gamma_n$, $\beta$, and $\Delta$(0) (specific heat) were obtained for sample No.\,2, $\Delta$(0) (AR) -- for sample No.\,1. Interval of $2\Delta(0)/T_c$ include parameters for both samples.}
		\end{table}

		In summary, we performed comprehensive  study of the band structure and superconducting properties of the SnAs binary compound. 
		DFT/GGA band structure calculations were carried out for both bulk and (111) slab crystal structures  with and without SOC. 
		Our calculated spectra for bulk SnAs are in a good agreement with  previous theoretical results. 
		Experimental investigation of SnAs band structure was done by ARPES technique. 
		Comparison of the ARPES results and band structure calculation shows that the calculated (111) slab band structure much better agrees with ARPES data than the bulk band structure. 
		SOC does not have a strong influence on electronic structure of (111) slab. 
		From our data we deduced quantitative parameters and the  type of the superconducting state in SnAs. 
		More specifically, \\
		(i)  we confirmed the type I superconductivity in SnAs by features of the specific heat in non-zero magnetic fields, and by the relationship between the estimated quantitative superconducting parameters; \\
		(ii) we found that the temperature dependences of critical magnetic field,  specific heat jump, and superconducting energy gap are consistent with the conventional weak-coupling BCS model. 
		Our data shows no signature of the unconventional  superconductivity, and the superconducting state in SnAs is likely to have  the $s$-type symmetry;\\
		(iii) using two independent techniques (specific heat and Andreev reflection spectroscopy) we determined the superconducting energy gap value $\Delta(0) = 0.6$\,meV - from specific heat data,  and $\Delta(0) = 0.53 \pm 0.04$\,meV - from Andreev reflection spectroscopy. 
		These values also satisfy the weak coupling BCS relationship $2\Delta(0)/T_c$=3.52. 
		
		Table \ref{tab:parameter} summaries our obtained values of the SnAs parameters and compare them with earlier data.

	\section{Acknowledgements}
		
		 Authors thank for valuable discussions E.Z. Kuchinskii and M.V. Sadovskii.
		 
		This work was performed using equipment of the LPI Shared Facility Center and  the resource center ``Physical methods of surface	investigation'' (PMSI) of the Research park of Saint Petersburg State University. 
		
		VMP acknowledges RFBR grant \#16-29-03330. 
		PIB, KAD, AVS, KSP, AVM, ASU, AYT, SYG works were supported within the state assignment of the Ministry of Science and Higher Education of the Russian Federation (project  No. 0023-2019-0005).
		
		AGR acknowledge St. Petersburg State University for research Grant No. 15.61.202.2015.
		
		IAN, AAS and NSP aknowledge RFBR grants \#17-02-00015 and \#19-32-50001, the Program No. 12 of Fundamental Research of the  Presidium of RAS. 
		NSP work was also supported in part by the President of Russia grant for young scientists No. MK-1683.2019.2.

	%

\end{document}